\documentstyle[aps,prl,floats,psfig]{revtex}

\begin{document}

\twocolumn[{
\draft
\title{Resonant Spheres: Multifrequency Detectors of Gravitational Waves}

\author{M. Angeles Serrano\cite{angelesmail} and J. Alberto Lobo\cite{albertomail}}

\address{
         Departament de F\'\i sica Fonamental, Universitat de Barcelona,
         Diagonal 647, 08028 Barcelona, Spain}
 
\date{ \today }

\maketitle
\widetext
\begin{abstract}
\leftskip 54.8 pt
\rightskip 54.8 pt
We discuss the capabilities of spherical antenn\ae\/ as single 
multifrequency detectors of gravitational waves. A first order theory allows us to evaluate the coupled spectrum
and the resonators readouts when the first and the second quadrupole bare sphere frequencies 
are simultaneously selected for layout tuning. We stress the existence of
non-tuning influences in the second mode coupling causing draggs in the 
frequency splittings. These URF effects are relevant to a correct physical
description of resonant spheres, still more if operating as multifrequency
appliances like our PHCA proposal.  
\end{abstract}
\leftskip 54.8 pt


}]

\narrowtext

A spherical antenna is particularly well adapted to sense metric tidal 
gravitational wave 
excitations as a consequence of the perfect matching between the GW 
radiation patterns and the sphere's vibration eigenmodes. When suitably 
monitored, it can act as an individual multimode detector generating potential 
knowledge on the incident direction of the signal, its quadrupole or monopole 
amplitudes, or its polarizations \cite{0DL,1DL}.

Spherical geometry also offers another substantial ability which optimizes the
capabilities of resonant bars currently in operation \cite{2DL}. By comparison, on equality 
of building material and mass, the estimated sphere's absorption cross 
section 
for the fundamental mode is over a factor of 4 better than that of a 
Weber's cylinder in its most favorable orientation. In addition, the second
sphere's quadrupole still shows a rather large cross section, only about half its 
value at the lowest frequency and still about 15 times bigger than that 
associated to the cylinder for the fundamental mode, which is the only 
showing a significant cross section in bars \cite{3DL}. 
Hence,{\it spheres show good sensitivity 
not only at the first but also at the second quadrupole harmonic, and therefore 
can efficiently operate at two frequencies as single multifrequency detectors.}
 
The challenge becomes how to take advantage of this potentiality. A preliminary 
task before multifrequency implementation design is to acquire a clear 
understanding of the detector's performance when working at one resonant 
frequency --see for instance \cite{9DL} or \cite{4DL}.

The theories are proposed for practical multimode devices consisting of a perfect
elastic sphere of radius $R$, mass $M$, density $\varrho$ and Lam\'e coefficients 
$\lambda$ and $\mu$, monitored by a suitable readout system,
commonly a set of $J$ identical 
resonators modelled as linear springs of mass $m \equiv \eta M, \eta << 1$, and endowed
at certain positions ${\bf n}_a \equiv {\bf x}_a/R$ on the sphere's surface. Each 
resonator  
amplifies the elastic radial deformation $u_a(t)$ of the sphere at its attachment 
point caused by an external force and supplies an essential increase
in the coupling by measuring the springs' deformations $q_a(t)$ relative
to the sphere's undeformed surface. Resonant tuning requires the sensors to be  
typically built to possess a natural frequency $\Omega$ equal to
one of the eigenfrequencies
of the free sphere's spectrum, say $\omega_{NL}$. This is always taken to be
a monopole 
or a quadrupole frequency --$\omega_{N0}$ or $\omega_{N2}$ respectively--, since only 
monopole or quadrupole sphere's spheroidal modes can be excited by an 
incoming wave \cite{5DL}. 

In the universally assumed ideal approximation, the tuning frequency $\omega_{NL}$ 
is thought to be
an isolated resonance frequency (IRF): no other frequency $\omega_{nl}$ of the
free sphere's spectrum interferes in the resonance. For instance, this assumption
is correct when the resonators couple to $\omega_{12}$, in practice the most interesting
coupling for unifrequency quadrupole radiation sensing \cite{3DL}. 

We follow the perturbative approach in \cite{4DL} naturally opened to refined analysis, 
and recall that the responses of the coupled device to an incident GW
can be inferred from the system

\begin{eqnarray}
   \varrho \frac{\partial^2  u_a}{\partial t^2} - \mu\nabla^2 u_a
       - (\lambda+\mu)\,\nabla(\nabla{\bf\cdot} u_a)  =  {\bf f}_{res} +
       {\bf f}_{GW} \hspace{0.8cm}\nonumber \\
    \ddot{q}_{a}(t)  +  \Omega_{a}^2\,q_{a}(t)  = - \ddot{u}_{a}(t) + \zeta_{a,GW}(t)  
      \qquad a=1,\ldots,J , 
\end{eqnarray}
where ${\bf f}_{res}(x_a,t)$ and ${\bf f}_{GW}(x_a,t)$ are respectively due to the 
resonators' back action on the sphere and to the incident GW which also
causes a tidal acceleration $\zeta_{a,GW}(t)$ on resonator $a$.

After implementation of Green function formalism, the equations can be rewriten
in the $s$-Laplace domain as

\begin{eqnarray}
\sum_{b=1}^J M_{ab} \,\hat q_b(s) = -\frac{s^2}{s^2+\omega_{NL}^2}\,
  \hat u_{a,GW}(s) + \frac{\hat\zeta_{a,GW}(s)}{s^2+\omega_{NL}^2}\ , \nonumber \\
   a=1,\ldots,J ,
\end{eqnarray}
with $\hat u_{a,GW}(s)$ the bare sphere's radial
responses to the signal at the resonators' locations. 

In the IRF circumstance, matrix $M_{ab}$ is of the form

\begin{equation}
   M_{ab} \equiv \left[\,\delta_{ab} + \eta\,\frac{s^2\omega_{NL}^2}{s^2+
   \omega_{NL}^2}\,P_{L,ab}\,\right] .
\end{equation} 

The geometric properties of a particular resonator arrangement are displayed by the 
$J\times J$ symmetric real matrix $P_L$ associated to the $L$-mode selected for 
tuning. It basically has as element $ab$ the Legendre 
polynomial of order $L$, a sum of products of spherical harmonics: 

\begin{equation}
P_{L,ab} \,\, = \,\, |A_{NL}(R)|^2 \,\,\sum_{m=-L}^{m=L} Y_{LM}^*({\bf n}_a)Y_{LM}({\bf n}_b)
\end{equation}
with $A_{NL}(R)$ radial components in the spheroidal normal modes of the free sphere \cite{5DL}. 
It is highly remarkable that all the information determining the distinctive readout of a given 
configuration is just concentrated the eigenvalues $\xi_{a}^{2}$ and eigenvectors $v^{(a)}$ 
of $P_L$.

First order resolution for the spectrum of coupled-mode eigenfrequencies 
from $\det M_{ab} = 0 $
shows that the attachment of resonators causes the
tuning frequency $\omega_{NL}$ to split into $J$ {\it symmetric pairs} around the original
value,

\begin{equation}
\omega_{a,\pm}^2 \,\,=\,\, \omega_{NL}^2 \, \left( 1 \pm \,\xi_{a}
               \eta^{\frac{1}{2}} \right) + O(\eta) \, \qquad a=1,\ldots,J ,
\end{equation}
whereas resonators amplitudes amount to be linear combinations of the GW amplitudes
$\hat{g}^{(2m)}(s)$:

\begin{eqnarray}  
\hat{q}_a(s) = \eta^{-\frac{1}{2}} \sum_{b=1}^J \sum_{\pm \xi_{c} \neq 0} \left\{ 
F_{L}(\pm \xi_{c}) 
 \frac{1}{(s^2 + \omega_{c,\pm}^2)} v_a^{(c)}v_b^{(c)*} \right\}\times 
 \hspace{-0.5cm} \nonumber \\
\times \left. \sum_{m=-2}^{m=2}Y_{2m}({\bf n}_b) \hat{g}^{(2m)}(s) \right\}+ O(0) 
    \qquad a=1,\ldots,J.\\
F_{L}(\pm \xi_{c}) \,= \, \pm\left[a_{NL} A_{NL}(R)\right] \, (-1)^J (\xi_{c})^{-1} \, \hspace{2cm}
 \nonumber 
\end{eqnarray}
with $a_{NL}$ non zero overlap coefficients only if $L=0$ or $L=2$\cite{5DL}. 

\begin{table}[htb]
\begin{tabular}{|c|c|c|} \hline && \\
Layout&\multicolumn{2}{c|}{IRF Eigenfrequencies}  \\
&around $\omega_{12}$& around $\omega_{22}$\\ \hline \hline
&& \\
& $\omega_{0,\pm}=\omega_{12}(1\pm 0.58\eta^{\frac{1}{2}})$ &
 $\omega_{0,\pm}=\omega_{22}(1\pm 0.03\eta^{\frac{1}{2}})$ \\
&& \\
PHCA & $\omega_{1,\pm}=\omega_{12}(1\pm 0.88\eta^{\frac{1}{2}})$&
$\omega_{1,\pm}=\omega_{22}(1\pm 0.05\eta^{\frac{1}{2}})$ \\ 
&&\\
& $\omega_{2,\pm}=\omega_{12}(1\pm 1.07\eta^{\frac{1}{2}})$&
$\omega_{2,\pm}=\omega_{22}(1\pm 0.06\eta^{\frac{1}{2}})$ \\  
&&  \\ \hline \hline 
& &  \\
TIGA & $\omega_{\pm}=\omega_{12}(1\pm 1.00 \eta^{\frac{1}{2}})$&
$\omega_{\pm}=\omega_{22}(1\pm 0.05 \eta^{\frac{1}{2}})$ \\  
& &\\ \hline
\end{tabular}
\caption{IRF frequencies around $\omega_{12}$ and $\omega_{22}$ for the PHCA 
and TIGA arrangements.}
\end{table}

However, this is not always the case that the interesting 
frequency for tuning is an
isolated resonant frequency. For example, the second quadrupole $\omega_{22}$
is in fact a suitable sphere's frequency for a second resonator set to be tuned to in order
to exploit the antenna as a multifrequency device. A careful examination of the spectrum
of a typical planned full scale aluminium 
sphere --$\eta_s \approx 1/40000$, $R=1.5m$-- around 
$\omega_{22}$ shows that $\omega_{14}$ is
merely within a distance of order $\eta^{\frac{1}{2}}$:

\begin{equation}
\omega_{14}^2  \,\,= \,\, \omega_{22}^2 \, \left( 1 + K \,\,\,
\eta^{\frac{1}{2}} \, \right) ,
\end{equation}
where the dimensionless parameter $K$ takes the negative value $K=-2.21$ for $\eta=\eta_s$.

The expectation is that this nearness alters in some way the
above IRF results. The effects of such unisolation --URF effects \cite{6DL}--
must be accurately considered to reach a faithful description of the detector's real
behaviour.

Let us analyse the general case when $\omega_{NL}$ is the single selected frequency 
for tuning and $\omega_{N'L'}$ is in its neighbourhood. Again, we start from 
expressions (1) and (2) which remains unchanged, although a new 
contribution appears in (3): 

\begin{eqnarray}
   M_{ab} \equiv \left[\,\delta_{ab} \,\,+ \,\,\eta\,\frac{s^2\omega_{NL}^2}{(s^2+ 
   \omega_{NL}^2)^2}\,P_{L,ab}\,\right.+
   \nonumber \\
   + \left. \eta\,\frac{s^2\omega_{N2}^2}{(s^2+\omega_{N2}^2)(s^2+ \omega_{N'L'}^2)}\,
    P_{L',ab}\,\right] .
\end{eqnarray} 

We can still go further in drawing generic conclusions valid for any resonator 
distribution whenever it allows $P_L$ and $P_{L'}$ to commute: $[P_L,P_{L'}] =0$.
Then, we name $v^{(a)}$ the orthogonal basis which simmultaneously diagonalises
the two matrices, with eigenvalues $\xi^2_{a,L}$ and $\xi^2_{a,L'}$ respectively,
so that in this basis $M_{ab}$ is also diagonal. Then, the URF resonances around 
$\omega_{NL}$ can be written in $J$ {\it triplets}

\begin{eqnarray}
\omega_{a,T}^2 \,\,=\,\, \omega_{NL}^2 \, \left( 1 + 
U_{a}^{T} \eta^{\frac{1}{2}} \right) + O(\eta) \hspace{1cm}\nonumber \\
 a=1,\ldots,J ,
\end{eqnarray}
where for each resonator index $a$, the upper label $\{T\}$ represents the group  
$\{u,c,d\}$ which refers to the three different 
solutions of the cubic equation

\begin{eqnarray}
U_{a}^3 \,\,- \,\,K \cdot U_{a}^2\,\, -\,\,\chi^2 \cdot U_{a}\,
+\, K\,\xi^2_{a,2}\,\,=\,\,0 \nonumber \\
\chi^2 \,\, = \,\,\xi^2_{a,2}\,\,+ \,\,\xi^2_{a,L'}, 
\end{eqnarray}
with parameters which are univocally determined by fixing the layout, $\omega_{NL}$ 
and $\omega_{N'L'}$. Inspection of orders of magnitude in (10) for $\omega_{NL}=\omega_{22}$ and 
$\omega_{N'L'}=\omega_{14}$ leads to the
conclusion that the triplets present a general pattern with independence of the 
resonator distribution: one of the three frequencies will always be located very close
to the original tuning frequency $\omega_{22}$ --to assess how much near this frequency
actually is one needs to restrict to particular cases and numerical evaluations--. The
remaining pair forms a non-symmetric doublet around it, in good agreement with the idea that
the presence of $\omega_{14}$ causes a perturbation of the ideal IRF situation. 

Then, the URF effect results in a dragging effect breaking the symmetry of the IRF 
doublets which approach the disturbing frequency $\omega_{14}$, and moreover induces 
the appearance of a third central component near the resonant $\omega_{22}$.

Near actually means really near, at least for two known proposals: PHCA \cite{7DL,8DL} 
and TIGA \cite{9DL,10DL}. 
It can be seen from the numbers in Table II that in each a-group the central $U_a^c$ 
is itself 
of order $\eta^{\frac{1}{2}}$ or smaller, so that these central URF resonances actually differ 
from $\omega_{22}$ in terms of order $\eta$. Reproduction of the amplitudes 
evaluation process demostrates that the contribution of  
such modes are not at leading order $\eta^{-\frac{1}{2}}$
but at higher terms. Therefore,  they are referred to as be
weakly coupled. 
\begin{table}[htb]
\hspace{-4cm}\begin{tabular}{|c|c|c|} \hline 
& & \\
Layout&Strongly Coupled&Weakly Coupled \\
& &  \\ \hline \hline
& &  \\
& $\omega_{0,u}=\omega_{22}(1+0.06\eta^{\frac{1}{2}})$& $\omega_{0,c}=$\\ 
& $\omega_{0,d}=\omega_{22}(1-1.16 \eta^{\frac{1}{2}})$&
$\omega_{22}(1-3.3\cdot 10^{-3} \eta^{\frac{1}{2}})$ \\
& &  \\
PHCA & $\omega_{1,u}=\omega_{22}(1+1.41\eta^{\frac{1}{2}})$&$\omega_{1,c}=$   \\ 
&$\omega_{1,d}=\omega_{22}(1-2.51 \eta^{\frac{1}{2}})$&
$\omega_{22}(1-1.5\cdot 10^{-4} \eta^{\frac{1}{2}})$ \\ 
& &  \\
& $\omega_{2,u}=\omega_{22}(1+1.07\eta^{\frac{1}{2}})$ &  $\omega_{2,c}=$\\  
& $\omega_{2,d}=\omega_{22}(1-2.18\eta^{\frac{1}{2}})$ &
$\omega_{22}(1-3.5\cdot 10^{-4}\eta^{\frac{1}{2}})$  \\
& &  \\ \hline \hline 
& &  \\
& $\omega_{1-5,u}=\omega_{22}\,(1+1.22\eta^{\frac{1}{2}})$ & $\omega_{1-5,c}=$\\ 
TIGA & $\omega_{1-5,d}=\omega_{22}(1-2.32\eta^{\frac{1}{2}})$& 
$\omega_{22}(1-2.5\cdot 10^{-4}\eta^{\frac{1}{2}})$ \\ 
& &  \\ 
& $\omega_{6,S}=\omega_{22}(1-1.105\eta^{\frac{1}{2}})$ &
$\omega_{6,D}=\omega_{22}$ \\ 
& &  \\ \hline
\end{tabular}
\caption{SCD+WCS URF triplets around $\omega_{22}$. Calculation have been 
performed for the proposals PHCA and TIGA, and for 
$\eta_{s}$,a theoretical value for a full scale future 
spherical detector. Subindex $u$ (up) labels the 
values which are above the resonance frequency $\omega_{22}$, whereas $d$ (down) labels
those underneath, and $c$ (central) those practically at $\omega_{22}$.}
\end{table}

The result is that both PHCA and TIGA URF triplets are 
composed of a a weakly coupled singlet plus a strongly coupled
doublet (triplet named of the SCD+WCS type) of span comparable to 
that of the IRF pairs.The only exception is the sixth mode in TIGA 
which corresponds to a triplet composed of 
a strongly coupled singlet plus a weakly coupled doublet (SCS+WCD type). Also, the degeneracy 
pattern of the IRF triplets is maintained: for PHCA, three different triplets, two of them doubly
degenerated, and for TIGA one triplet five-fold degenerate plus one single weakly coupled triplet. 

As said, only strongly coupled frequencies (SC) contribute to the amplitudes:

\begin{eqnarray} 
 \hat{q}_a(s) = \eta^{-\frac{1}{2}} \sum_{b=1}^J \sum_{SC} \left\{ 
F_{LL'}(U_{c}^{SC}) 
 \frac{1}{(s^2 + \omega_{c,SC}^2)} v_a^{(c)}v_b^{(c)*} \right\}
 \hspace{-0.9cm}
 \nonumber \\
 \times  \sum_{m=-2}^{m=2}Y_{2m}({\bf n}_b) \hat{g}^{(2m)}(s)+ O(0)  
 \!\!\!\qquad a=1,\ldots,J,\hspace{0.2cm} 
\end{eqnarray}
with frequency weights

\begin{eqnarray} 
F_{LL'}(U_{c}^{SC})= \left[ a_{NL} A_{NL} + a_{N'L'} A_{N'L'} \frac{U_{c}^{SC}}
 {U_{c}^{SC}-K}\right] \times \hspace{-0.4cm}\nonumber \\
 \times\frac{U_{c}^{SC}-K}{\displaystyle{\prod_{SC \neq SC' \neq SC''}} 
 (U_{c}^{SC'} - U_{c}^{SC})(U_{c}^{SC''} - U_{c}^{SC})} .
 \hspace{0.6cm}
\end{eqnarray}

The high degree of symmetry showed by both PHCA and TIGA explains why these 
configurations fulfil the property $[P_L,P_{L'}]=0$ for $L=2$ and $L'=4$, so that $P_2$ and $P_4$
simultaneously diagonalise. After algebraic calculus, the eigenvectors of the common
basis are finally found to be arrangements of spherical harmonics of order two, precisely the 
eigenvectors of $P_2$ when diagonalised separately in the IRF situation:

\begin{equation}
v_a^{(m)} \,\,=\,\, \sqrt{\frac{4\, \pi}{5}} \xi_{m,(2)}^{-1}\, Y_{2m}({\bf n}_a).
\end{equation}
By inserting these functions in (12), we arrive to simplified expressions for the readouts 
of the PHCA layout:

\begin{eqnarray} 
\hat{q}_{a,PHCA}(s) \,\,=\,\,\eta^{-\frac{1}{2}}  \sum_{SC}
\left\{  \,F_{24}(U_{0}^{SC})\,\,
  \frac{ Y_{20}({\bf n}_a) \hat{g}^{(20)}(s)}{(s^2 + \omega_{0,SC}^2)}\right.
  \nonumber \hspace{-0.5cm}\\
 +\,\, F_{24}(U_{1}^{SC})\,\,
 \frac{\left[ Y_{21}({\bf n}_a) \hat{g}^{(21)}(s)+Y_{2-1}({\bf n}_a) 
 \hat{g}^{(2-1)}(s)\right] }{(s^2 + \omega_{1,SC}^2)} 
  \nonumber \\
+\left.\,\, F_{24}(U_{2}^{SC})\,\,
  \frac{\left[ Y_{22}({\bf n}_a) \hat{g}^{(22)}(s)+Y_{2-2}({\bf n}_a) 
  \hat{g}^{(2-2)}(s)\right]}{(s^2 + \omega_{2,SC}^2)}  \right\}; \nonumber \hspace{-0.5cm}\\
\end{eqnarray}
and the readouts of the TIGA layout:
\begin{equation} 
\hat{q}_{a,TIGA}(s)  = \eta^{-\frac{1}{2}}  \sum_{SC}
 F_{24}(U_{1-5}^{SC})\,\,
\frac{\displaystyle{\sum_{m=-2}^{m=2}}Y_{2m}({\bf n}_a) \hat{g}^{(2m)}(s)}
{(s^2 + \omega_{1-5,SC}^2)}. 
\end{equation}
The weight functions $F_{24}(U_{c})^{SC}$ in (14) and (15) can be settled from (12).
It is here important to recall that the overlap coefficient $a_{14}$ is zero, 
which is in accordance with the idea that only monopole or quadrupole sphere's modes
can be excited by an impinging GW. One has to take also into account that
just $up$ and $down$ frequencies in the 
triplets correspond to SC values, and that the
the remaining $U_{c}^{c}$ give place to
weakly coupled resonances which do not contribute at leading order. Hence,

\begin{equation}
F_{24}(U_{c}^{SC})\,=\,F_{24}(U_{c}^{u,d})\,=\,
a_{22}A_{22}\,\frac{1}{U_{c}^{d,u}-U_{c}^{u,d}}\,\frac{U_{c}^{d,u}}{U_{c}^{u,d}} .
\end{equation}

In Figure 1 we display numerical simulations for the IRF and URF qualitative
responses caused by the action 
of a GW burst travelling down the symmetry axis of the PHCA pentagonal layout.
In both cases the practical outputs become beats, although differing in
modulation and modulated frequencies. The IRF output exactly cancels at the nodes
because the
wheights affecting the contributions to the amplitudes of $\omega_+$ and $\omega_-$
equal each other
whereas the URF readout shows a certain thickness at those same points caused by
the difference between the wheights associated to $\omega_u$ and $\omega_d$. The
occurrence of different wheights is in direct relation to the dragging 
induced by the URF effect which breaks the symmetry of the IRF-doublets.
Nevertheless, the similarity of the patterns again corroborates the weakness 
of the new third central frequency in the URF triplets.

\vspace{-6.5cm}
\begin{figure}[hbt]
\centerline{\hspace{-0.8cm}
\psfig{file=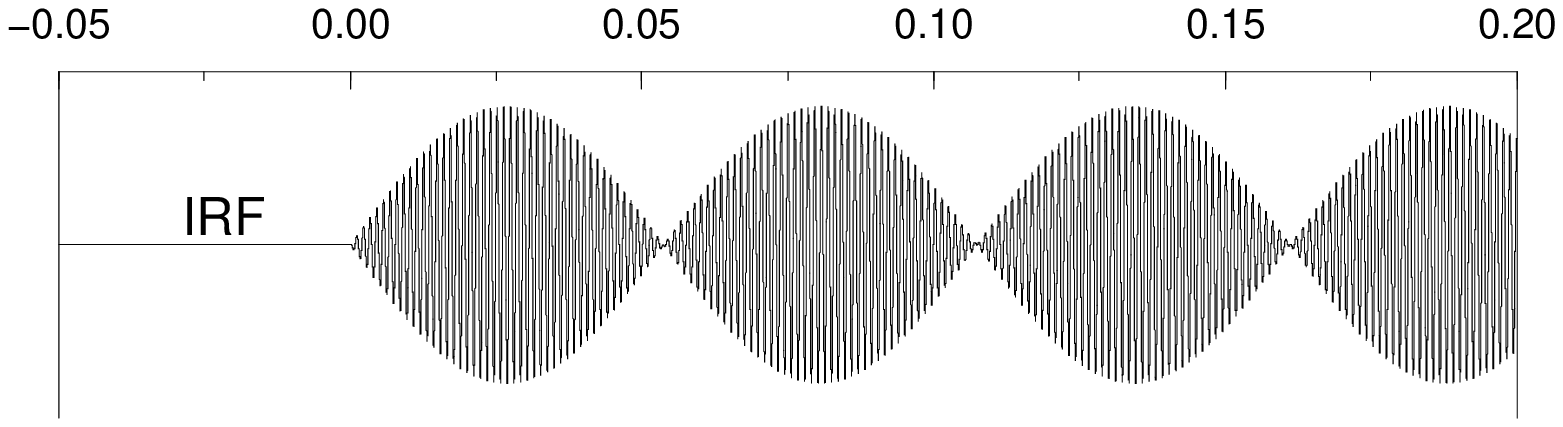,height=11cm,width=11cm}}
\vspace{-9cm} 
\centerline{\hspace{-0.8cm}
\vspace{-1cm}
\psfig{file=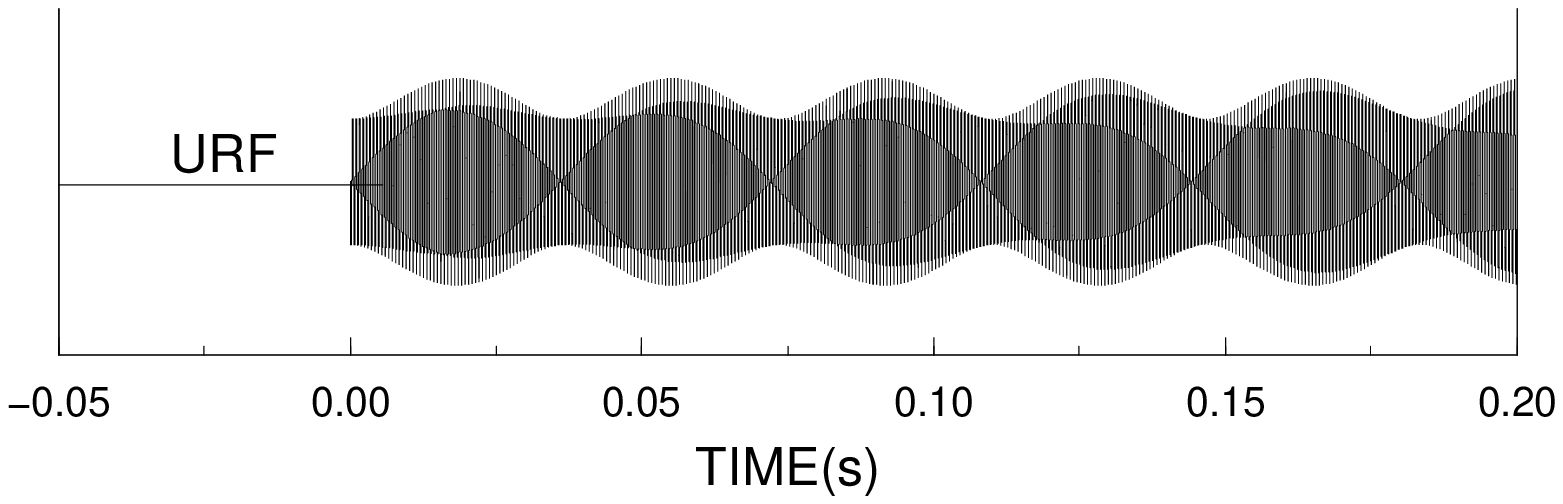,height=11cm,width=11cm}} \vspace{0.3cm}
\caption{PHCA beat outputs for GW burst with $h_{+}=h_{\times}$ propagating
along the pentagonal symmetry axis at t=0.}
\end{figure}

Hence, the analysis of the URF effect is essential for a complete study of any spherical 
multifrequency GW detector, but we are specially interested in it with respect to our PHCA 
proposal. 
\begin{figure}[htb]
\vspace*{-0.6cm}
\hspace{-1.1cm}
\centerline{
\psfig{file=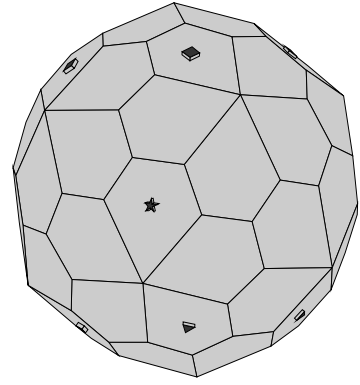,height=7.5cm,width=4.8cm}}\vspace*{-1.1cm}
\caption{PHCA Proposal. Marked faces in the pentagonal hexacontahedron indicate 
resonator positions: a {\it square\/} for transducers tuned to $\omega_{12}$, 
a {\it triangle\/} for that tuned to $\omega_{22}$, and a {\it star\/} for the 
monopole sensor.}
\end{figure}

It is
actually idealized as a multifrequency antenna with two sets of supplementary pentagonal 
layouts, one tuned to the first quadrupole harmonic $\omega_{12}$ and another coupled to 
the second quadrupole harmonic
$\omega_{22}$, which in fact has $\omega_{14}$ very close to it. The pentagonal hexacontahedron
polyhedric shape with more than enough 
number of faces for 10 resonators matching in non parallel pentagonal
positions (and even an eleventh resonator position for monopole sensing) guarantees technical 
feasibility, also  supported by the absence of
cross interactions between the outputs of each set at order $\eta^{-\frac{1}{2}}$.  
Our model ensures a correct
description of the multifrequencial PHCA coupled spectrum. We find two main 
groups of frequencies, the first composed of pairs symmetrically distributed
around $\omega_{12}$ and the second arising as a non-symmetrical
splitting of $\omega_{22}$ in triplets dragged towards $\omega_{14}$. 

\begin{figure}[htb]
\hspace{2.2cm}
\centerline{\psfig{file=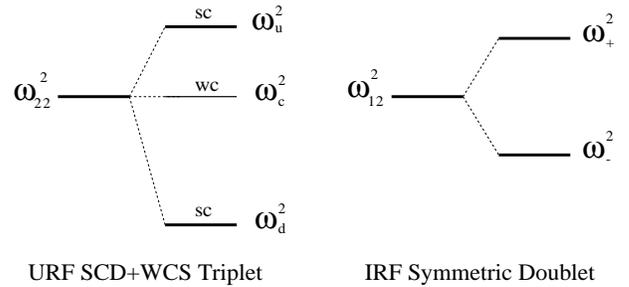,height=3.8cm,width=8cm}}
\vspace{0.1cm}
\caption{PHCA multifrequency spectrum.}
\end{figure}

The PHCA monitored at their first two quadrupole modes can for instance be 
advantageously used to detect chirp signals from 
coalescencing binary systems, and even to determine some of their 
characteristic parameters by means of a robust double passage method \cite{11DL}.
This potentiality, unthinkable for currently operating bars, shows that resonant 
spherical antennae are abreast of projected broadband large laser interferometers 
with respect to their predicted ability in monitoring gravitational waves from these
sources \cite{12DL}.

In any case, the URF effect plays {\it per se} an essential part 
in a rigorous theoretical 
analysis of any single multifrequency GW detector. Its features are concisely reported 
by our developments without severely complicating the evaluations. 
We conclude by emphasizing that the philosophy underlying these algorithms 
is easily extensible to the study of other practical 
real situations departing from ideality. 

M. A. Serrano thanks Tom\'as Alarc\'on and Miquel Montero for active encouragement, and 
Jos\'e M. Pozo for many helpful mathematical discussions.
This research is supported by the Spanish Ministry of Education under grant
No. FP97-46725209 and contract No. PB96-0384, and the Institut d'Estudis Catalans.

\end{document}